\documentclass[preprint,twocolumn,3p]{elsarticle}

\usepackage{lineno,hyperref}
\modulolinenumbers[5]

\journal{Journal of \LaTeX\ Templates}

\begin{document}

\begin{frontmatter}

\title{Dependence of Vortex States in Superconductors on a Chiral Helimagnet and an Applied Magnetic Field}

\author[math]{Saoto Fukui \corref{Fukui}}
\ead{st110035@edu.osakafu-u.ac.jp}
%\address{Department of Mathematical Sciences, Osaka Prefecture University,1-1,Gakuencho,Naka-ku,Sakai,Osaka 599-8531,Japan.}

\author[math]{Masaru Kato}

\author[electronics]{Yoshihiko Togawa}
\cortext[Fukui]{Corresponding author. Address: Department of Mathematical Sciences, Osaka Prefecture University, 1-1, Gakuencho, Naka-ku, Sakai, Osaka 599-8531, Japan. 
Tel.:+81 72 254 9368; fax: +81 72 254 9916}

\address[math]{Department of Mathematical Sciences, Osaka Prefecture University, 1-1, Gakuencho, Naka-ku, Sakai, Osaka 599-8531,Japan}
\address[electronics]{Department of Physics and Elecrtronics, Osaka Prefecture University, 1-1, Gakuencho, Naka-ku, Sakai, Osaka 599-8531,Japan}

\begin{abstract}
We study effects of a chiral helimagnet (CHM) on vortex states in a superconductor,
solving the Ginzburg-Landau equations in a chiral helimagnet/superconductor bilayer system.
We found that vortices form a periodically modulated triangular lattice,
because the magnetic field from the chiral helimagnet$H_{\rm CHM}$ oscillates spatially.
An increase of a critical current is expected, because vortices are pinned by the $H_{\rm CHM}$.
\end{abstract}


\end{frontmatter}

%\linenumbers

\section{Introduction}
A vortex state is one of important features for applications of superconductors.
Vortex states depend on an external magnetic field and an external current.
It is known that a magnetic field from a ferromagnet affects the vortex state of the superconductor \cite{F/S_Review}.
For example, a ferromagnet / superconductor hybrid structure enhances superconductivity, especially its critical magnetic field and its critical current \cite{F/S_dot,F/S_bilayer}.

Magnetic materials like a ferromagnet affect a superconductor.
One of magnetic materials that attracts attention recently is a chiral helimagnet(CHM).
The CHM has an interesting magnetic structure. 
The CHM consists of spins that form the helical rotation along one direction.
A spin structure of the CHM comes from two interactions between nearest neighbor spins; a ferromagnetic exchange interaction and the Dzyaloshinsky-Moriya (DM) interaction \cite{Dzyaloshinsky}.
Two nearest neighbor spins tend to be parallel due to the former interaction but tend to be perpendicular to each other due to the latter interaction.
Due to a competition between two interactions, nearest neighbor spins slightly deviate from each other. This leads to a formation of the helically rotated arrangement.
When a weak magnetic field is applied, the magnetic structure forms a soliton lattice.
These magnetic structures have been observed experimentally \cite{Togawa_CSL}.

In the previous study, we found that the chiral helimagnet affects a vortex configuration in a superconductor \cite{ISS2014}.
Vortices form a periodically modulated triangular lattice under an applied large magnetic field.

In this paper, we study effects of a chiral helimagnet on a vortex configuration in a superconductor in more detail.
We compare vortex configurations with/without the magnetic field from the CHM under the applied homogeneous magnetic field.
In order to investigate vortex configurations, we solve the Ginzburg-Landau equations with the finite element method.
\begin{figure}[t]
  \begin{center}
   \includegraphics[scale=0.4]{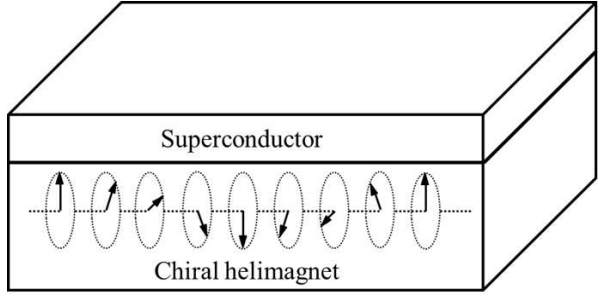}
   \caption{The chiral helimagnet/ superconductor bilayer system.}
 \label{Fig1}
   \end{center}
\end{figure}

\section{Method}
We consider a chiral helimagnet/superconductor bilayer system in Fig.\ref{Fig1}.

We assume that the effect of the chiral helimagnet on the superconductor is given by an external magnetic field $H_{\rm CHM}$ but effects of the superconductor on the chiral helimagnet are neglected.
In this study, the superconducting layer is considered as a two-dimensional system and only the perpendicular component of the magnetic field $H_{\rm CHM}$ is taken into account.

\begin{figure}
  \begin{center}
   \includegraphics[scale=0.4]{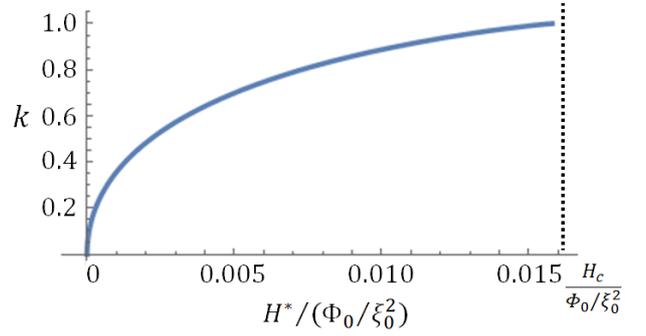}
   \caption{The relation between the modulus of the Jacobi's elliptic function $k$
    and the applied magnetic field $H_{\rm appl}$ for $D/J = 0.16$}
 \label{Fig2}
   \end{center}
\end{figure}

In order to investigate vortex configurations in this bilayer system, we solve the Ginzburg-Landau equations,
\begin{equation}
  \alpha |\psi|^2 + \beta |\psi|^2 \psi + \frac{1}{2m^\ast} 
  \left( \frac{\hbar}{i}\nabla - \frac{e^\ast}{c}\mbox{\boldmath $A$} \right)\psi = 0,  \label{gl-1} 
\end{equation}
\begin{eqnarray}
 && {\rm curl}\left( {\rm curl}\mbox{\boldmath $A$} - \mbox{\boldmath $H$}_{\rm ext} \right) = \frac{4\pi}{c}\mbox{\boldmath $J$} \nonumber  \\
 && = \frac{4\pi}{c}\left\{ \frac{e^\ast \hbar}{2m^\ast i} \left( \psi^\ast \nabla \psi - \psi \nabla \psi^\ast \right) \right. \nonumber \\
 && \left. - \frac{e^{\ast2}}{m^\ast c} \psi^\ast \psi \mbox{\boldmath $A$} \right\}, \label{gl-2} 
\end{eqnarray}
where $\alpha=\alpha_0(T-T_c)$, $T$ is a temperature, $T_c$ is a critical temperature, 
$\alpha_0 (>0)$ and $\beta (>0)$ are coefficients, $\psi$ is an superconducting order parameter, 
$m^\ast$ is an effective mass, $e^\ast$ is an effective charge, $\mbox{\boldmath $A$}$ is a magnetic vector potential, 
$\mbox{\boldmath $H$}_{\rm ext}$ is an external magnetic field, and $\mbox{\boldmath $J$}$ is a supercurrent density.
In the second equation, the Maxwell equation is included.
The magnetic field from the chiral helimagnet $H_{\rm CHM}$ is included in the external magnetic field $H_{\rm ext}$.
$H_{\rm CHM}$ is obtained from a Hamiltonian for chiral helimagnet \cite{kishine}.
This Hamiltonian consists of a ferromagnetic exchange interaction, the Dzyaloshinsky-Moriya interaction, 
and the Zeeman energy;
\begin{eqnarray}
 \mathcal{H} &=& -J \sum_n \mbox{\boldmath $S$}_n \cdot \mbox{\boldmath $S$}_{n+1} + \mbox{\boldmath $D$} \cdot \sum_n \mbox{\boldmath $S$}_n \times \mbox{\boldmath $S$}_{n+1} \nonumber \\
             & & + 2\mu_BH_z \sum_n \mbox{\boldmath $S$}_n^z, \label{hamiltonian}
\end{eqnarray}
where $\mbox{\boldmath $S$}$ is a spin in the chiral helimagnet, $J$ is an exchange coefficient, $\mbox{\boldmath $D$}$ is a DM vector.
From this Hamiltonian, we obtain the perpendicular component of the magnetic field $(\mbox{\boldmath $H$}_{\rm ext})_z$;
\begin{equation}
 \left(\mbox{\boldmath $H$}_{{\rm ext}}\right)_z (x) = H_0 \cos{\theta} + H_{{\rm appl}}.  \label{external_field} 
\end{equation}
The first term is a magnetic field from the chiral helimagnet $H_{\rm CHM}$, where
\begin{equation}
 \theta = 2 {\rm sin}^{-1} \left[{\rm sn} \left( \frac{\sqrt{H^\ast}}{k}x | k \right) \right] + \pi.
\end{equation}
$H_0$ is strength of the magnetic field from the CHM.
The second term is an applied magnetic field $H_{\rm appl}$.
$H^\ast$ is a normalized magnetic field,
\begin{equation}
 H^\ast = \frac{2\mu_B H_{\rm appl}}{a^2 S^2 \sqrt{J^2+D^2}}, \label{norm_H}
\end{equation}
where $a$ is a lattice constant.
$k$ $(0 \leq k \leq 1)$ is a modulus of the Jacobi's elliptic function ${\rm sn}(x|k)$ and determined by,
\begin{equation}
 \frac{\pi \phi}{4\sqrt{H}^\ast} = \frac{E(k)}{k}. \label{k}
\end{equation}
$\phi = \tan^{-1}{\left(D/J \right)}$ and $E(k)$ is the complete elliptic integral of the second kind.
A relation between $k$ and $H^\ast$ is shown in Fig.\ref{Fig2}.

\begin{figure}
  \begin{center}
  \includegraphics[scale=0.33]{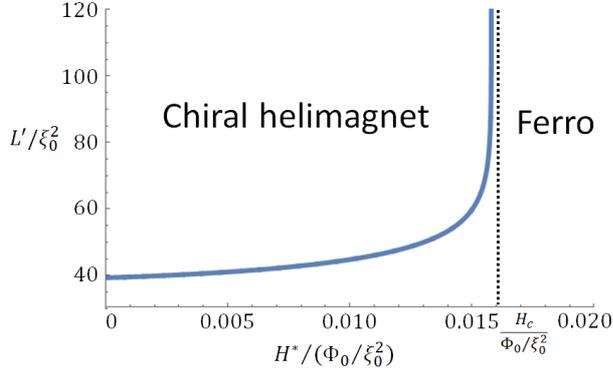}
  \caption{The relation between the helical period and the applied magnetic field for $D/J = 0.16$.}
  \label{Fig3}
  \end{center} 
\end{figure}

A period of the helical rotation $L'$ is given by,
\begin{equation}
 L' = \frac{2kK(k)}{\sqrt{H^\ast}}, \label{period}
\end{equation}
where $K(k)$ is the complete elliptic integral of the first kind.
A relation between $L$ and $H^\ast$ is shown in Fig.\ref{Fig3}.

We obtain stable states using the Ginzburg-Landau equations, which is solved by the finite element method \cite{d-dot}.

\section{Result}
We show vortex configurations in a two-dimensional superconductor system under the $H_{\rm CHM}$ and $H_{\rm appl}$.
We take the Ginzburg-Landau parameter $\kappa = \lambda_0/\xi_0=10$ ($\lambda_0$ and $\xi_0$ are a penetration length and coherence length at $T=0$, respectively), the temperature $T=0.3T_c$, 
and the ratio between two interactions $D/J = 0.16$, 
which is taken from the experimental data for Cr$_{1/3}$NbS$_2$ \cite{D/J}.
The system sizes are $5.0L'\xi_0 \times 40\xi_0$.
We take boundary conditions: (a) $\mbox{\boldmath $A$} \cdot \mbox{\boldmath $n$} = 0$,where $\mbox{\boldmath $n$}$ is a normal vector to the surface, (b) edges in this system are free.

\begin{figure}
 \begin{center}
 \includegraphics[scale=0.5]{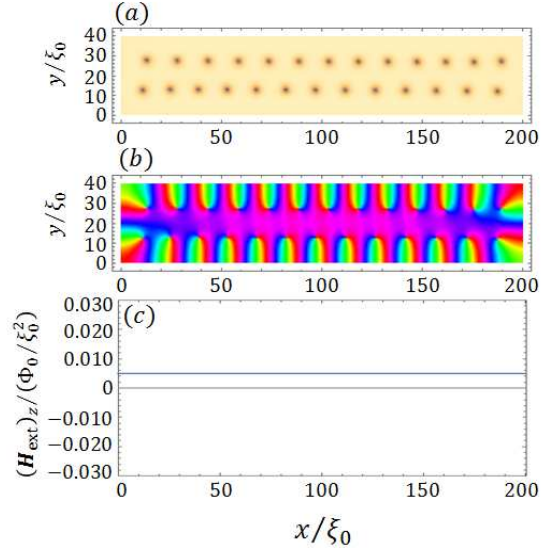}
 \caption{(a) Distributions of order parameters, (b) phases, and (c) magnetic fields,
          for $H_0/(\Phi_0/\xi_0^2)=0.000$ and $H_{{\rm appl}}/(\Phi_0/\xi_0^2)=0.0050$.}
 \label{Fig4}
 \end{center}
\end{figure}
\begin{figure}
 \begin{center}
 \includegraphics[scale=0.5]{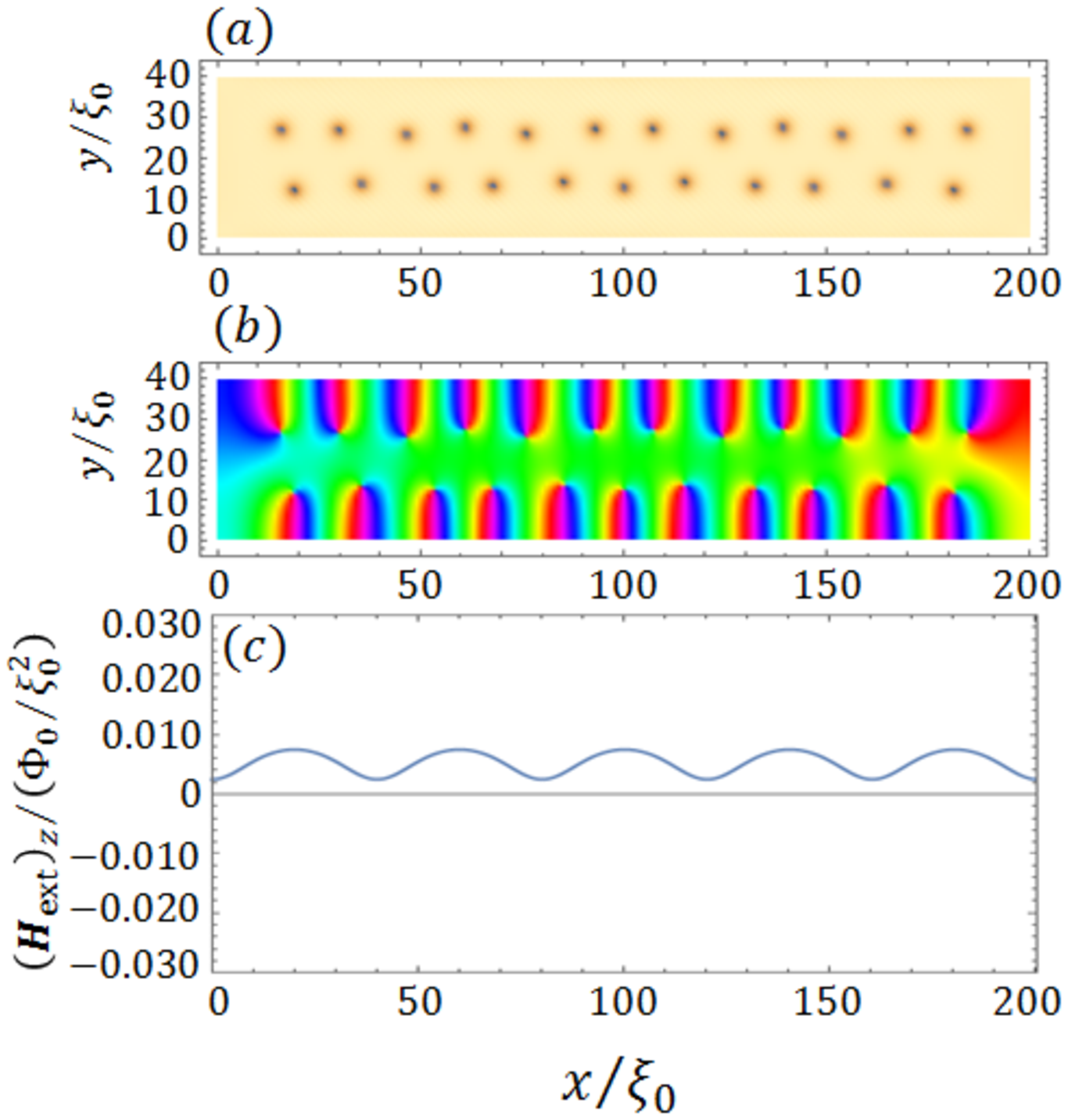}
 \caption{(a) Distributions of order parameters, (b) phases, and (c) magnetic fields,
          for $H_0/(\Phi_0/\xi_0^2)=0.0025$ and $H_{{\rm appl}}/(\Phi_0/\xi_0^2)=0.0050$.}
 \label{Fig5}
 \end{center}
\end{figure}
\begin{figure}
 \begin{center}
 \includegraphics[scale=0.5]{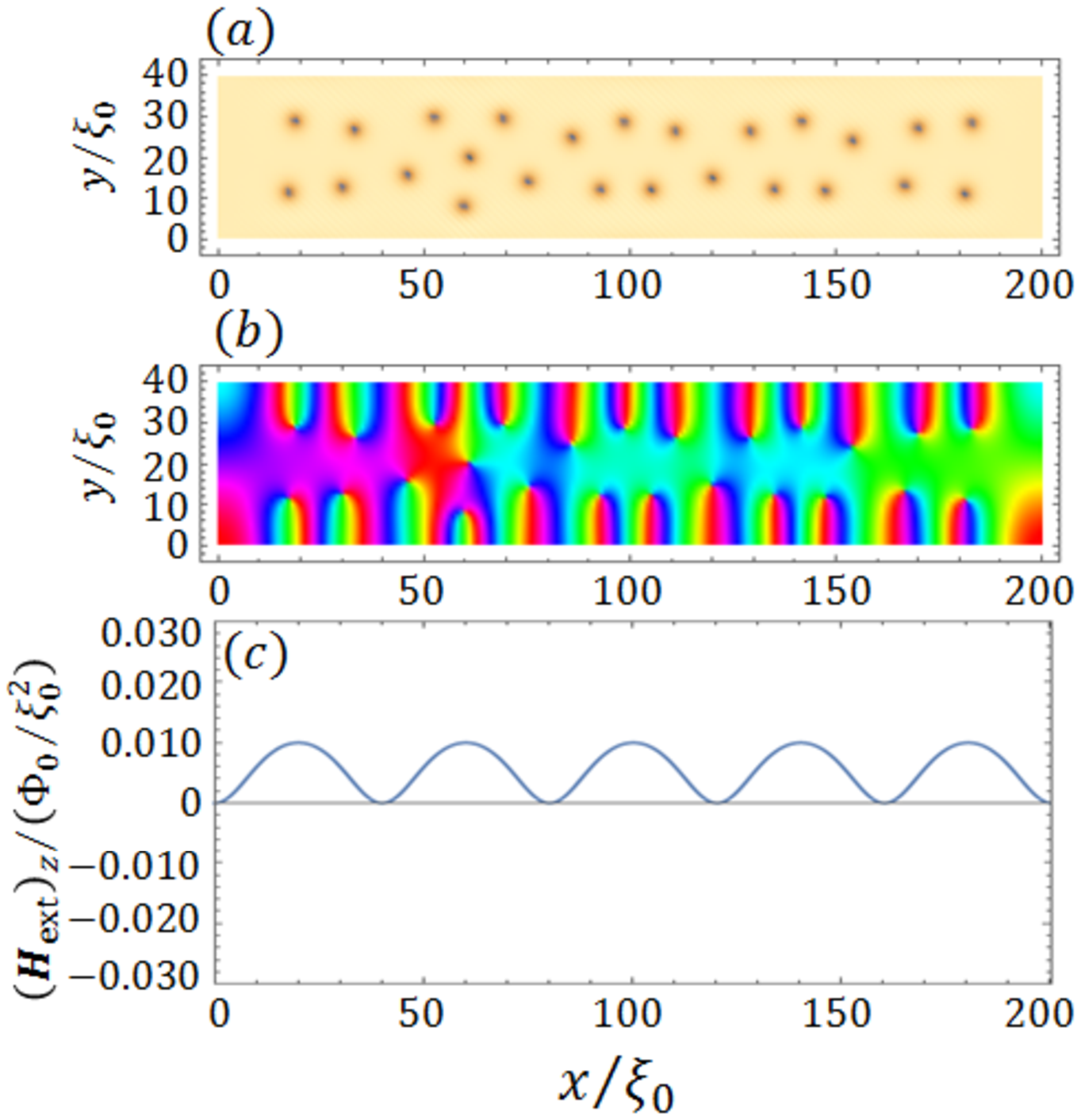}
 \caption{(a) Distributions of order parameters, (b) phases, and (c) magnetic fields,
          for $H_0/(\Phi_0/\xi_0^2)=0.0050$ and $H_{{\rm appl}}/(\Phi_0/\xi_0^2)=0.0050$.}
 \label{Fig6}
 \end{center}
\end{figure}
\begin{figure}
 \begin{center}
 \includegraphics[scale=0.5]{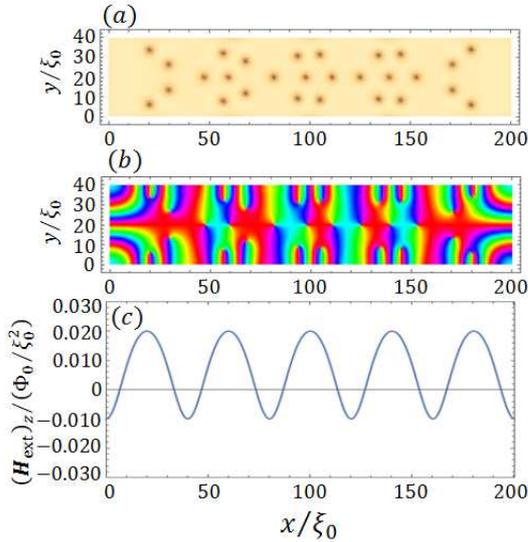}
 \caption{(a) Distributions of order parameters, (b) phases, and (c) magnetic fields,
          for $H_0/(\Phi_0/\xi_0^2)=0.0150$ and $H_{{\rm appl}}/(\Phi_0/\xi_0^2)=0.0050$.}
 \label{Fig7}
 \end{center}
\end{figure}

We show vortex configurations under the applied magnetic field in Fig.\ref{Fig4}-\ref{Fig7} for $H_{\rm appl}/(\Phi_0/\xi_0^2) = 0.0050$, where $\Phi_0$ is an quantum flux.
In Figs.\ref{Fig4} (a)-(c), their figures represent a distribution of the order parameter, the phase, and the magnetic field for $H_0=0.0$ and $H_{\rm appl}/(\Phi_0/\xi_0^2) =0.0050$.
In the case Figs.\ref{Fig4}, the Abrikosov lattice forms.
When the magnetic field from the chiral helimagnet $H_{\rm CHM}$ is given, vortex configurations change to Fig.\ref{Fig5}-\ref{Fig7}, where $H_0/(\Phi_0/\xi_0^2) $ are $0.0025$ (Fig.\ref{Fig5}), $0.0050$ (Fig.\ref{Fig6}), and $0.0150$ (Fig.\ref{Fig7}) and $H_{\rm appl}/(\Phi_0/\xi_0^2) $ are fixed to $0.0050$.
Under $H_{\rm CHM}$, triangular lattices are modulated. 

We can explain this result by a following reason;
this modulation comes from the magnetic field from the chiral helimagnet $H_{\rm CHM}$.
The chiral helimagnet has a helical magnetic structure, so $z$-component of the magnetic field $(\mbox{\boldmath $H$}_{\rm CHM})_z$ oscillates spatially.
For $H_0/(\Phi_0/\xi_0^2) = 0.0025$ (Fig.\ref{Fig5}), $H_{\rm CHM}/(\Phi_0/\xi_0^2)$ changes from $-0.0025$ to $0.0025$ and $H_{\rm ext}/(\Phi_0/\xi_0^2) =H_{\rm CHM}/(\Phi_0/\xi_0^2)  + H_{\rm appl}/(\Phi_0/\xi_0^2) $ changes from $0.0025$ to $0.0075$ for $H_{\rm appl}/(\Phi_0/\xi_0^2) = 0.0050$.
Vortices tend to appear in the large magnetic field region and avoid in the small magnetic field region due to an interaction between the vortex and the magnetic field.
The latter interaction $E_{VF}$ is given by,
\begin{equation}
 E_{VF} = -\frac{1}{4\pi} \mbox{\boldmath $\Phi$}_0 \cdot \mbox{\boldmath $H$}_{\rm ext}. \label{vh_interaction}
\end{equation}
Therefore, a periodic modulated triangular lattice is formed.

Here, we discuss a dynamics of a vortex under the applied magnetic field and the magnetic field from the chiral helimagnet.
When a current flows along the $y$-direction, vortices move along the $x$-direction.
However, under $H_{\rm CHM}$, vortices is difficult to move through the region where the magnetic field is low or negative.
Therefore, a movement of a vortex is restricted and a pinning effect of the vortex due to the chiral helimagnet is expected.
This leads to the increase of the critical current.

\section{Summary}
We have investigated the effect of the chiral helimagnet on the vortex configuration.
Under the applied magnetic field and the magnetic field from the CHM, vortices form the periodically modulated triangular lattice.
It is expected that this vortex configuration leads to a pinning effect and the increase of the critical current.

\section*{Acknowledgements}
This work was supported by JPSJ KAKENHI Grant Number 26400367.
The authors thank V.V. Moshchalkov, Y. Kato, T. Nojima, T. Ishida, S. Okuma, S. Mori, Y. Ishii, Y. Higashi, N. Fujita, M. Umeda, M. Kashiwagi for useful discussions.

\end{document}